\documentstyle[12pt,epsfig]{article}
\begin{document}
\begin{titlepage}
\vspace{1.0cm}
\begin{center}
{\large \bf {CHIRAL SU(3) DYNAMICS WITH COUPLED CHANNELS:\\ 

\smallskip
INCLUSION OF P-WAVE MULTIPOLES }}

\vspace{1.0cm}

{\large J. Caro Ramon, N. Kaiser, S. Wetzel and W. Weise}

\vspace {0.8cm}
Physik-Department, Technische Universit\"{a}t M\"{u}nchen\\
   Institut f\"{u}r Theoretische Physik, D-85747 Garching, Germany

\end{center}
\vspace{1.0cm}

\begin{abstract}
We extend our recent non-perturbative chiral SU(3) coupled channel approach to
pion- and photon-induced $\eta$- and $K$-meson production off protons by 
including all strong and electromagnetic p-wave multipoles. We identify the 
p-wave amplitudes of the next-to-leading order SU(3) chiral meson-baryon 
Lagrangian with a coupled channel potential which is iterated to infinite 
orders in a separable Lippmann-Schwinger equation. Our approach to $\eta$- and
$K$-photoproduction introduces no additional free parameters. By adjusting a
few finite range parameters and the unknown parameters in the Lagrangian, we
are able to simultaneously describe a very large amount of low-energy data.
These include the total and differential cross sections of the $\pi$-induced
reactions  $\pi^- p \to \eta n, K^0 \Lambda, K^0 \Sigma^0, K^+ \Sigma^-$ and 
$\pi^+ p \to K^+ \Sigma^+$ as well as those of photoproduction $\gamma p \to
\eta p , K^+\Lambda, K^+ \Sigma^0, K^0 \Sigma^+$. The  polarization 
observables measured in $\eta$- and $K$-photoproduction are particularly 
sensitive to interference terms between the s- and p-wave multipoles.
The total cross section data are remarkably well reproduced in all
channels. There remain, however, some open questions concerning details of 
angular distributions and polarization observables.  

\end{abstract}

\vspace{2in}

Accepted for publication in {\it Nuclear Physics} {\bf A}.
\vfill
\noindent $^*${\it Work supported in part by BMBF and DFG}
\end{titlepage}
\vskip 2.5cm

\section{Introduction}

In a recent publication \cite{kww} we have developed a theoretical model to 
study  simultaneously pion- and photon-induced production of eta- and K-mesons
from protons, based on our previous work in ref.\cite{siegel}. In contrast to 
most other approaches \cite{adelsack,mart,sauermann} 
we do not introduce explicit (baryonic or mesonic) resonances. Our starting 
point is the SU(3) chiral effective Lagrangian for the octet baryons ($N,
\Lambda,\Sigma,\Xi$) and the octet pseudoscalar Goldstone bosons ($\pi, K,\eta
$) with interactions controlled by chiral symmetry and a low-momentum
expansion. In ref.\cite{kww} we have identified the s-wave meson-baryon 
amplitudes generated by the SU(3) chiral Lagrangian with a coupled channel
potential matrix which is  iterated to infinite order in a separable 
Lippmann-Schwinger equation in momentum space. The resulting multi-channel 
S-matrix is automatically unitary, and one naturally obtains cusp effects at
secondary thresholds (e.g. at the $K\Sigma$-threshold for reactions with
$K\Lambda$-final states).  Furthermore, strong attraction is predicted by 
the SU(3) chiral Lagrangian in certain s-wave channels, namely $\bar K N$  
with total isospin 0 and  $K\Sigma$ with total isospin 1/2. As a consequence, 
quasi-bound s-wave meson-baryon states emerge in the coupled channel scheme. In
fact these states possess all the properties attributed to the baryon 
resonances $\Lambda(1405)$ and $S_{11}(1535)$ \cite{siegel}. In particular the
unstable $K\Sigma$-bound state has a large branching ratio for decaying into
$\eta N$ which is the outstanding feature of the nucleon resonance
$S_{11}(1535)$.    

Remaining adjustable parameters (a priori unknown coefficients in the 
Lagrangian and short range cut-offs in the separable Lippmann-Schwinger 
equation) are fixed in a simultaneous fit to a large amount of low energy data.
These include the total cross sections for elastic and
inelastic $K^-p$-scattering ($K^-p \to K^-p,\bar K^0n, \pi^0\Lambda,
\pi^0\Sigma^0, \pi^-\Sigma^+, \pi^+ \Sigma^-$) as well as the total cross
sections for pion- and photon-induced $\eta$- and $K$-production ($\pi^-p
\to \eta n, K^0 \Lambda, K^0 \Sigma^0, K^+ \Sigma^-$ and $\gamma p \to \eta p,
K^+ \Lambda, K^+ \Sigma^0, K^0 \Sigma^+$). Among these data the most precise 
ones are those for $\gamma p \to \eta p$ measured  at MAMI \cite{krusche}. One
finds that the combined MAMI/ELSA $\eta$-photoproduction total cross sections
\cite{krusche,schoch} can be well reproduced in s-wave approximation. On the
other hand, for the  kaon photoproduction channels $\gamma p \to K^+ \Lambda,
K^+ \Sigma^0$  measured at ELSA \cite{bockhorst,goers}, the s-wave
approximation works only in the  
near threshold region. At higher energies s-waves alone considerably 
underestimate the total cross sections and evidently p-waves start to become 
important. This is also seen from the angular distributions of the differential
cross section which are no longer isotropic at these energies. Furthermore, the
$\Lambda$- and $\Sigma$-recoil polarizations measured at ELSA \cite{bockhorst}
(via the self-analyzing weak decay of these hyperons) can result only from s- 
and p-wave interference terms. Also, existing differential cross sections for  
pion-induced $\eta$- and  $K$-production point toward the importance of
the meson-baryon p-wave amplitudes. 

The purpose of this work is to extend our previous coupled channel framework 
\cite{kww,siegel} by including, in a systematic manner, the strong interaction 
$p_{1/2}$- and $p_{3/2}$-partial wave amplitudes together with the 
photoproduction p-wave multipoles ($E_{1+},M_{1+}, M_{1-})$. The latter
correspond to the electric quadrupole transition ($E_{1+}$) and the magnetic 
dipole transitions ($M_{1\pm}$) in pseudoscalar meson photoproduction. In 
complete analogy with the treatment of the s-waves in ref.\cite{kww} we will 
identify the (strong and electromagnetic) p-wave multipoles given by  
the first and second order SU(3) chiral meson-baryon Lagrangian with a coupled
channel potential. This matrix of driving terms is iterated to infinite orders
in a separable (p-wave) Lippmann-Schwinger equation using the same values of
the short range cut-offs as for the s-waves. For the strong meson-baryon
interaction seven new parameters come along with four spin-non-flip terms and
three spin-flip terms at next-to-leading order. 

A very welcome feature of our approach is that the 
inclusion of the electromagnetic interaction ($\eta$ and $K$-photoproduction) 
is completely parameterfree. The respective p-wave multipoles ($E_{1+},M_{1+},
M_{1-})$ depend only on well known baryon magnetic moments and the SU(3) axial
vector coupling constant $F$ and $D$. It is then quite non-trivial to achieve 
a good description of so many empirical photon- and pion-induced data
within a coupled channel approach, based on a single chiral effective
Lagrangian, in which all these processes are correlated.

In this paper we do not aim for an accurate reproduction of elastic $\pi
N$-phase shifts. This is a drawback of the present approach which could be
fixed by parametrizing the relevant $\pi N$-amplitudes. Our goal here is rather
to focus on the impact of chiral $SU(3)_L\times SU(3)_R$ on hadron dynamics
involving kaons and $\eta$-mesons. The more detailed influence of the elastic
$\pi N$-channel will be investigated at a later stage.   

On the experimental side there are now improved data (total and differential
cross sections and hyperon recoil polarizations) for kaon photoproduction 
from ELSA \cite{goers,prelim}. In the case of $\eta$-photoproduction the 
polarized photon asymmetry has recently been measured at GRAAL \cite{didelez},
and the polarized target asymmetry was measured at ELSA \cite{bock}. At
Brookhaven there is an ongoing program \cite{cbc} to measure $\pi^- p \to \eta
n$ and related reactions.    

This paper is organized as follows. In section 2 we present the formalism to
calculate the strong and electromagnetic p-wave amplitudes within our approach.
We first display the relevant p-wave part of the SU(3) chiral meson-baryon
Lagrangian at second order, derive expressions for the strong p-wave 
potentials and formulate the separable p-wave Lippmann-Schwinger equation. 
Next we apply the same formalism to $\eta$- and $K$-photoproduction and give 
expressions for the observables (total and differential cross sections as well 
as the polarized target asymmetry, the recoil polarization and the polarized
photon asymmetry) in s- and p-wave approximation. In section 3 we discuss in 
detail our results for pion- and photon-induced  $\eta$- and
$K$-meson production off protons, and we conclude in section 4 with a summary.

\section{Formalism}
\subsection{Chiral effective Lagrangian}
The leading order relativistic SU(3) chiral meson-baryon Lagrangian and the
s-wave part at next-to-leading order have already been developed and explained
in detail in section 2.1 of ref.\cite{kww}. Therefore we display here only the
new terms relevant to p-wave meson-baryon scattering at second order in small
momenta \cite{guido}:  
\begin{eqnarray} {\cal L}_{\phi B}^{(2)} &=& 2g_D\, {\rm tr} (\overline B \{ 
\vec u \cdot \vec u, B\} ) + 2g_F\, {\rm tr} (\overline B [\vec u \cdot \vec u
,B]) +2g_0 \, {\rm tr}(\overline BB)\,{\rm tr} (\vec u \cdot \vec u) \nonumber
\\ & &+ 2g_1\, {\rm tr} (\overline B \vec u) \cdot {\rm tr}(\vec u B)
+2 h_D\, {\rm tr} (\overline B i\vec \sigma \cdot \{ \vec u \times \vec u, B\}
) + 2h_F\, {\rm tr} (\overline B i\vec \sigma \cdot [\vec u \times \vec u
,B]) \nonumber \\ & &+ 2h_1\, {\rm tr} (\overline B i\vec \sigma \times \vec u)
\cdot {\rm tr}(\vec u B). \end{eqnarray}
Here, the SU(3)-matrix $B$ represents the octet baryons ($N,\Lambda,\Sigma,\Xi
$). The axial vector quantity $u^\mu = -\partial^\mu \phi/2f$ is proportional
to  the gradient of the octet pseudoscalar meson fields ($\pi, K,\eta$)
represented by the SU(3)-matrix $\phi$ and $f=92.4$ MeV denotes the pion decay
constant, $\vec \sigma$ is the usual spin-vector given by the $2\times 2$
Pauli-matrices. Note that in flavor SU(3) there are four independent 
non-spin-flip terms with coefficients $g_D, g_F, g_0, g_1$ and three 
independent spin-flip terms with coefficients $h_D, h_F, h_0$. Further 
possible  SU(3)-symmetric  p-wave interaction terms at this order can be 
expressed through the ones given in eq.(1) using non-trivial trace relations
\cite{guido}. 

When computing the photoproduction p-wave multipoles at a later stage, we will
need the photon-baryon vertex at next-to-leading order, i.e. the terms
proportional to baryon anomalous magnetic moments. In SU(3) the corresponding
part of the  Lagrangian reads
\begin{equation} {\cal L}^{(2)}_{\gamma B}= {1\over 2M_0} {\rm tr} \Big( 
\overline B \vec \sigma \cdot (\vec \nabla \times \vec A) ( \kappa_D \{ Q,B\} +
\kappa_F [Q,B])\Big)\,, \end{equation} 
where $\vec A$ denotes the usual electromagnetic vector potential, $Q={1\over 
3}{\rm diag}(2,-1,-1)$ is the quark charge matrix and $M_0 \simeq 0.9$ GeV the 
octet baryon mass in the chiral limit \cite{bugra}. In the chiral limit (of 
vanishing quark masses $m_u=m_d=m_s=0$) the baryon anomalous magnetic moments 
are related to the two constants $\kappa_D$ and $\kappa_F$ in eq.(2) as
follows,   
\begin{eqnarray} &&\kappa_p={1\over 3} \kappa_D+\kappa_F=1.79, \quad  \kappa_n 
= -{2\over 3} \kappa_D=-1.91, \quad \kappa_\Lambda = -{1\over 3} \kappa_D 
=-0.61, \nonumber \\ && \kappa_{\Sigma^+} = {1\over 3} \kappa_D+\kappa_F=1.46, 
\quad \kappa_{\Sigma^0} = {1\over 3} \kappa_D=0.65, \quad \kappa_{\Lambda 
\Sigma^0} = {1\over \sqrt 3} \kappa_D=1.61, \end{eqnarray}
However, as the empirical values given in eq.(3) indicate, there are 
significant deviations from the simple SU(3)-relations. To be more accurate at
this point, we will actually use the empirical values of the baryon anomalous 
magnetic moments inasmuch as they are available. The empirically unknown 
$\kappa_{\Sigma^0}$ and the sign of the transition magnetic moment 
$\kappa_{\Lambda\Sigma^0}$ are taken in eq.(3) from the
recent chiral  perturbation theory calculation of ref.\cite{sven}.  This
completes the discussion of the p-wave extension of the chiral effective 
Lagrangian. The seven unknown parameters $g_D, g_F, g_0, g_1, h_D, h_F,h_1$
will  be fixed in a fit to many scattering data in section 3.

\subsection{Coupled channel approach}
Let us now describe the coupled channel approach to the strong (p-wave) 
meson-baryon interaction. We label the meson-baryon states $|\pi N\rangle$, 
$|\eta N\rangle$, $|K \Lambda \rangle$, $|K \Sigma \rangle$ with total isospin
$I=1/2$ and the states $|\pi N\rangle$, $|K \Sigma \rangle$ with total isospin 
$I=3/2$ by an index which runs from 1 to 6, in that order. In a specific 
channel the meson center-of-mass energy  $E_\phi$ is given by      
\begin{equation} E_\phi = {s-M_B^2 + m_\phi^2 \over 2\sqrt{s}}\,,\end{equation}
where $\sqrt s$ is the total center-of-mass energy of the meson-baryon system
and $M_B, m_\phi$ denote the mass of the corresponding baryon and meson.
The center-of-mass momentum in channel $j$ reads 
\begin{equation} k_j = \sqrt{E_\phi^2 -m_\phi^2} = i \sqrt{m_\phi^2-E_\phi^2}\,
, \end{equation} 
and we have also given its analytical continuation below threshold (i.e. for 
$E_\phi < m_\phi$). 

Because of the spin-orbit interaction there are two different p-wave 
amplitudes in meson-baryon scattering (corresponding to the $p_{1/2}$- and 
$p_{3/2}$-states), and these do not mix. In
both angular momentum states the different channels are coupled through a 
(reduced) p-wave potential in momentum space,      
\begin{equation}
V_{ij}^{(1\mp)} = {\sqrt{M_i M_j}\over 4 \pi f^2 \sqrt{s}}\, C_{ij}^{(1\mp)}\,
\,, \end{equation}
where we have taken away the characteristic p-wave factor $k_i k_j$ which will
appear explicitly in the coupled channel equation. The 
relative coupling strengths $C_{ij}^{(1\mp)}$ are calculated from the chiral
SU(3) meson-baryon Lagrangian at next-to-leading order. Explicit expressions 
for these coupling strengths and a discussion of various 
contributions can be found  in appendix A . 

In the next step the potential $V_{ij}^{(1\mp)}$ is
iterated to all orders in a separable Lippmann-Schwinger equation of the form
\begin{equation} k_i \widetilde T_{ij}^{(1\mp)} k_j = k_i V_{ij}^{(1\mp)}k_j + 
\sum_n \int {d^3 l \over 2\pi^2} {1\over k_n^2 +i0 -l^2}\bigg( {\alpha_n^2
+k_n^2 \over \alpha_n^2 + l\,^2 } \bigg)^4 k_i V_{in}^{(1\mp)} l_n \, l_n
\widetilde T_{nj}^{(1\mp)} k_j \,. \end{equation}
The resulting multi-channel p-wave T-matrix is $T_{ij}^{(1\mp)} = k_i 
\widetilde T_{ij}^{(1\mp)} k_j$. The separable Lippmann-Schwinger equation can
be solved analytically by simple matrix inversion, 
\begin{equation} T_{ij}^{(1\pm)} = k_i \Big[ \Big( 1-V^{(1\pm)} \cdot
G^{(1)}\Big)^{-1} \cdot V^{(1\pm)} \Big]_{ij} k_j\,, \end{equation}
where $G^{(1)}$ is the diagonal matrix with elements
\begin{equation} G_n^{(1)} = {2\over \pi} \int_0^\infty{dl\,l^4 \over k_n^2 +i0
-l^2} \bigg( {\alpha_n^2+k_n^2  \over \alpha_n^2+l^2} \bigg)^4 = {k_n^2 -
\alpha_n^2 \over 16 \alpha_n^3} (\alpha_n^4 +10 \alpha_n^2 k_n^2 +k_n^4) - i\,
k_n^3 \,, \end{equation} 
and the center-of-mass momentum $k_n$ in the intermediate state $n$ is given 
by eq.(3) together with its analytical continuation below threshold. Note that
we are using here the same values for the short range cut-offs $\alpha_n$ as 
for the s-waves in ref.\cite{kww}. The exponent on the off-shell form factor
$[(\alpha_n^2+k_n^2)/(\alpha_n^2+l^2)]^4$ follows from the general rule $2L+2$
for given orbital angular momentum $L$ \cite{kloet}. The resulting 
multi-channel p-wave S-matrix $S_{ij}^{(1\mp)} = \delta_{ij} -2i
\sqrt{k_ik_j} \, T_{ij}^{(1\mp)}$ is exactly unitarity in the subspace of the 
kinematically open channels. 

Note that in the sense of chiral (i.e. small 
momentum) counting the iterations generated by the separable Lippmann-Schwinger
equation  do not affect the leading order current-algebra results for s-waves
\cite{kww,siegel}. In the p-wave channels, the formal chiral power counting is
distorted by the short distance cut-off scale $\alpha_n$ in eq.(7). It should 
be pointed out that the $KY$ and $\eta N$ channels investigated in this work 
have dynamic features quite different from those familiar from near threshold
$\pi N$ scattering. In particular there is no primary reason to expect that 
leading order current-algebra results would be relevant at the much higher
energy (around 1\,GeV) involved in $KY$- and $\eta N$-production. The main 
purpose of our investigation is to explore to what extent chiral $SU(3)_L\times
SU(3)_R$ still plays a significant role in constraining the driving terms
$V_{ij}^{(1\mp)}$ in the Lippmann-Schwinger equation eq.(7).    

In order to calculate observables for $\pi$-induced $\eta$- and $K$-meson
production one first has to relate the corresponding (s- and p-wave) T-matrix
elements to the ones defined in the isospin-basis. One finds for the physical
reaction channels,  
\begin{eqnarray} \pi^- p \to \eta n: & & \quad {\sqrt 6\over 3}
T_{12}^{(l\mp)}\,, \nonumber \\ \pi^- p \to K^0\Lambda: & & \quad 
{\sqrt 6\over 3} T_{13}^{(l\mp)}\,, \nonumber \\ \pi^- p \to K^0\Sigma^0: & &
\quad {\sqrt 2\over 3} \Big[ T_{56}^{(l\mp)} - T^{(l\mp)}_{14} \Big]
\,,\nonumber \\ \pi^- p \to K^+ \Sigma^-:  & & \quad {1\over 3} \Big[ 2T_{14}^
{(l\pm)}+ T^{(l\pm)}_{56} \Big]  \,,\nonumber \\ \pi^+ p \to K^+\Sigma^+: & &
\quad T_{56}^{(l\pm)} \,,\end{eqnarray}
and the superscript $(0+)$ refers to the s-wave amplitudes which have already
been calculated in ref.\cite{kww}. For each specific reaction channel the
differential cross section reads  
\begin{equation} {d\sigma\over d\Omega} = {k_{\rm out} \over k_{\rm in}} \Big\{
\big| T^{(0+)} + z \,\big( 2T^{(1+)}+T^{(1-)} \big)\big|^2 +(1-z^2) \big|
T^{(1+)}- T^{(1-)} \big|^2 \Big\}\,, \end{equation}
with $z= \cos \theta$ the cosine of the center-of-mass scattering angle 
$\theta$, and we have omitted the indices $i,j\in\{1,\dots 6\}$. The prefactor
$k_{\rm out}/k_{\rm in}$ is the usual flux and two-body phase-space
factor. Obviously, the angular distribution of the differential cross section
is restricted to a quadratic polynomial (of $z= \cos \theta$) in s- and p-wave
approximation. The total cross section for a specific reaction channel is
finally calculated as 
\begin{equation} \sigma_{\rm tot} = 4\pi {k_{\rm out } \over k_{\rm in}} \Big\{
\big|T^{(0+)}\big|^2 +2\big|T^{(1+)}\big|^2 +\big|T^{(1-)}\big|^2 \Big\} \,.
\end{equation} 

\subsection{Meson photoproduction}
We now extend the same formalism to p-wave meson photoproduction. As in
ref.\cite{kww} our basic assumption is that the photoproduction can be
described by a Lippmann-Schwinger equation. In complete analogy with the 
strong interaction part we identify the p-wave photoproduction
potentials (named generically $B_j$) with the electric quadrupole amplitude
$E_{1+}$ and the magnetic dipole amplitudes $M_{1\pm}$ calculated  from the
next-to-leading order chiral effective Lagrangian. For the description of the
photoproduction reaction $\gamma p \to \eta p, K^+\Lambda, K^+\Sigma^0, K^0
\Sigma^+$ we have to know the p-wave photoproduction potential $B_j$ for 
$\gamma p \to \phi B$, where $\phi B$ refers to the meson-baryon states with
total isospin $I=1/2$ or $I=3/2$ and isospin projection $I_3=+1/2$. In section
2.2 these states have been labeled by an index running from 1 to 6. Leaving out
the typical p-wave factor $k_j$ one finds for the reduced p-wave
photoproduction potentials,
\begin{eqnarray}
B_1 &=& {e M_N \over 8 \pi f \sqrt{3s}} ( D+F)\,\Big\{ 2 X_\pi(\kappa_n) +
Y_\pi(\kappa_p) \Big\} \,,\nonumber \\
B_2 &=& {e M_N \over 8 \pi f \sqrt{3s}}(3F-D)\,Y_\eta(\kappa_p)\,,\nonumber \\
B_3 &=& {e \sqrt{M_N M_\Lambda} \over 8 \pi f \sqrt{3s}}\bigg\{(-D-3F)X_K
(\kappa_\Lambda) + (F-D) {\kappa_{\Lambda \Sigma^0}\,\beta \over 2\sqrt 3 M_0}
\bigg\} \,,\nonumber \\ B_4 &=& {e \sqrt{M_NM_\Sigma} \over 8 \pi f \sqrt{3s}}
\bigg\{ ( D-F) \Big[ X_K(\kappa_{\Sigma^0}) + 2 Y_K(\kappa_{\Sigma^+}) \Big] +
(D+3F) {\kappa_{\Lambda\Sigma^0}\,\beta\over 6\sqrt3 M_0}\bigg\} \,,\nonumber 
\\ B_5 &=& {e M_N \over 4 \pi f \sqrt{6s}}( D+F)\Big\{ Y_\pi(\kappa_p)-X_\pi(
\kappa_n) \Big\} \,, \nonumber \\ B_6 &=&{e\sqrt{M_NM_\Sigma} \over 4 \pi f
\sqrt{6s}} \bigg\{ (D-F)\Big[ X_K(\kappa_{\Sigma^0})-Y_K(\kappa_{\Sigma^+})
\Big] + (D+3F) {\kappa_{\Lambda \Sigma^0}\, \beta \over 6 \sqrt 3 M_0} \bigg\} 
\,.\end{eqnarray}
These formulae apply separately to the three p-wave multipoles: $E_{1+},M_{1-}$
and $M_{1+}$. In these three cases the functions $Y_\phi(\kappa_+)$ (depending
on the anomalous magnetic moment $\kappa_+$ of a positively charged baryon) and
the functions $X_\phi(\kappa_0)$ (depending on the magnetic moment $\kappa_0$ 
of a neutral baryon) represent different expressions. If the photoproduced
pseudoscalar meson is neutral the functions $Y_\phi(\kappa_+)$ stand for   
\begin{eqnarray} Y_\phi[E_{1+}](\kappa_+) & = & 0 \,,\nonumber \\ 
Y_\phi[M_{1-}](\kappa_+) & = & -{4+3 \kappa_p+\kappa_+\over6M_0}\,,\nonumber \\
Y_\phi[M_{1+}](\kappa_+) & = & {1+\kappa_+ \over 3M_0} \,,\end{eqnarray}
and the argument in the square bracket indicates the p-wave multipole under
consideration. In the case that the photoproduced meson is positively charged 
the functions $X_\phi(\kappa_0)$ read  
\begin{eqnarray} X_\phi[E_{1+}](\kappa_0) & = & {1\over 12(E_\phi^2 -m_\phi^2)}
\bigg\{ E_\phi - {4 m_\phi^2 \over E_\phi} -{m_\phi^2 \over M_0} \Big( 1 +
{2m_\phi^2 \over E_\phi^2} \Big) + 3 \Big( 1 + {m_\phi^2 \over M_0 E_\phi}
\Big) L_\phi \bigg\} \,, \nonumber \\ 
X_\phi[M_{1-}](\kappa_0) & = & {1\over 2(E_\phi^2 - m_\phi^2) }\bigg\{ E_\phi
+{ m_\phi^2 -E_\phi^2 \over 3M_0} (3+3 \kappa_p +\kappa_0) - L_\phi \bigg\}\,,
\nonumber \\ X_\phi[M_{1+}](\kappa_0) & = & {1 \over 4(E_\phi^2 - m_\phi^2) }
\bigg\{ -E_\phi +{4\kappa_0 \over 3M_0} (E_\phi^2 - m_\phi^2) +L_\phi \bigg\} 
\,,\end{eqnarray}
with 
\begin{equation} L_\phi = {m_\phi^2 \over \sqrt{E_\phi^2 - m_\phi^2} } \ln
{E_\phi + \sqrt{E_\phi^2 -m_\phi^2} \over m_\phi} ={m_\phi^2 \over \sqrt{
m_\phi^2 -E_\phi^2} } \arccos {E_\phi  \over m_\phi}\,. \end{equation} 
The logarithmic function $L_\phi$ (together with its analytical continuation
below threshold) stems from the meson pole diagram in which the photon couples
to the positively charged meson (i.e. a $\pi^+$ or a $K^+$). In eq.(13) there
are further contributions from the electromagnetic $\Sigma^0 \to \Lambda
\gamma$ transition proportional to the transition magnetic moment
$\kappa_{\Lambda \Sigma^0}=1.61$. The corresponding weight factor $\beta$ is
different for the three p-wave multipoles and takes the values,
\begin{equation} \beta[E_{1+}] = 0\,, \quad \beta[M_{1-}] = 1\,, \quad 
\beta[M_{1+}] = -2\,. \end{equation}
This concludes the description of the p-wave photoproduction potentials derived
from the next-to-leading order SU(3) chiral Lagrangian. One finds that the
pertinent expressions depend only on the SU(3) axial vector coupling constant
$D\simeq 0.76$ and $F\simeq 0.50$  and known baryon magnetic moments. 

Infinitely many rescatterings of the photoproduced meson-baryon state due to
the p-wave strong interaction are summed up via the Lippmann-Schwinger
equation. The full p-wave photoproduction multipoles (commonly denoted by
${\cal M} \in \{ E_{1+},M_{1-},M_{1+}\}$) are then given by
\begin{equation} {\cal M}_i = k_i \sum_j \Big[ \Big(1-V^{(1\pm)} \cdot
G^{(1)} \Big)^{-1} \Big]_{ij} B_j \,,\end{equation}  
where the indices $i,j\in \{1,\dots,6\}$ refer to the isospin-basis. From
angular momentum conservation it is clear that the strong interaction 
$p_{3/2}$-potentials $V_{ij}^{(1+)}$ are the relevant ones for $E_{1+}$ and
$M_{1+}$, whereas $M_{1-}$ is iterated up by the $p_{1/2}$-potential
$V_{ij}^{(1-)}$. 

Next, we have to express the multipoles for the four physical photoproduction  
channels by the ones defined in the isospin-basis. One finds,  
\begin{eqnarray} \gamma p \to \eta p : & & \quad {\cal M}_2 \,,\nonumber \\ 
\gamma p  \to K^+\Lambda : & & \quad {\cal M}_3 \,, \nonumber \\ \gamma p \to 
K^+\Sigma^0 : & & \quad ( \sqrt2 {\cal M}_6 +{\cal M}_4 )/\sqrt 3 \,, 
\nonumber \\ \gamma p \to K^0 \Sigma^+ : & & \quad ( \sqrt 2 {\cal M}_4 -{\cal
M}_6 )/\sqrt 3\,. \end{eqnarray} 
In order to express the observables of meson photoproduction it is advantageous
to use the following linear combinations of three p-wave multipoles,
\begin{eqnarray} P_1 & = & 3E_{1+} +M_{1+}-M_{1-} \,, \nonumber \\ P_2 & = & 
3E_{1+} -M_{1+}+M_{1-} \,, \nonumber\\ P_3& =& 2M_{1+}+M_{1-}\,. \end{eqnarray}
In terms of these and the s-wave multipole $E_{0+}$ (see ref.\cite{kww}) the
differential cross section takes the form, 
\begin{equation} {d\sigma \over d\Omega} = {k_{\rm out} \over k_\gamma} \bigg\{
\big| E_{0+}+z \, P_1 \big|^2 +{1\over 2}(1-z^2) \Big( \big|P_2\big|^2+
\big|P_3\big|^2 \Big) \bigg\}\,, \end{equation}
with $k_\gamma = (s-M_N^2)/2\sqrt s$ the center-of-mass photon energy and
$z=\cos \theta$ the cosine of the center-of-mass scattering angle $\theta$. 
Again, the
angular distributions of the differential cross sections are restricted to a
quadratic polynomial (of $z=\cos \theta$) in s- and p-wave approximation. After
angular integration one gets the total meson photoproduction cross sections as
\begin{equation} \sigma_{\rm tot} = 4 \pi {k_{\rm out} \over k_\gamma} \Big\{
|E_{0+}|^2 +{1\over 3}\Big(|P_1|^2+|P_2|^2+|P_3|^2\Big)\Big\} \,.\end{equation}

Of particular interest are the polarization observables since they vanish
identically in pure s-wave approximation. The final state baryon emerges from
the photoproduction  process with a certain recoil polarization. This means 
that the probabilities for the baryon spin pointing downward or upward from 
the reaction plane spanned by the momenta of the photon and the meson are not
equal. In the case of kaon photoproduction the recoil polarization  $\Pi_r$ of
the hyperons $\Lambda, \Sigma^{0,+}$ is experimentally accessible through their
self-analyzing weak decays (into $\pi N$). In s- and p-wave approximation the
recoil polarization is calculated as follows from the multipoles: 
\begin{equation} \Pi_r ={ \sqrt{1-z^2}\, {\rm Im}[(E_{0+}+z\, P_1)(P_2+P_3)^*] 
\over |E_{0+}+z\,P_1|^2 +{1\over 2} (1-z^2) (|P_2|^2+|P_3|^2)}\,.\end{equation}
Another asymmetry can be measured by using a polarized proton target with the 
proton spin pointing either parallel or anti-parallel to the normal on the
reaction plane. In s- and p-wave approximation the polarized target asymmetry
reads
\begin{equation} A_t ={ \sqrt{1-z^2}\, {\rm Im}[(E_{0+}+z\, P_1)(P_3-P_2)^*] 
\over |E_{0+}+z\,P_1|^2 +{1\over 2} (1-z^2)(|P_2|^2+|P_3|^2) }\,.\end{equation}
Finally, one can produce the meson with linearly polarized photon beams whose 
polarization vector is either perpendicular or parallel to the reaction
plane. The corresponding polarized photon asymmetry reads in s- and p-wave
approximation 
\begin{equation} \Sigma_\gamma ={ (1-z^2)\, (|P_3|^2-|P_2|^2) 
\over 2|E_{0+}+z\,P_1|^2 +(1-z^2) (|P_2|^2+|P_3|^2) } \,. \end{equation} 
It is clear from eqs.(23,24,25), that even just the signs of the polarization
observables $\Pi_r, A_t, \Sigma_\gamma$ already contain non-trivial dynamical
information. 

\section{Results}
First, we have to fix parameters. As outlined in the previous section we are
working in the isospin-basis and we are using for the meson and baryon masses 
$m_\pi = 139.57$ MeV, $m_K= 493.65$ MeV, $m_\eta=547.45$ MeV, $M_N=938.27$ MeV,
$M_\Lambda= 1115.63$ MeV and $M_{\Sigma} = 1192.55$ MeV. This is a choice 
which averages out most isospin breaking effects. We keep the values of the 
axial vector coupling constants $D=0.782$, $F=0.485$, and those of the two 
chiral symmetry breaking parameters $b_D=0.066$ GeV$^{-1}$, $b_F=-0.231$
GeV$^{-1}$ unchanged from ref.\cite{kww}. In this work we identify  $M_0$ with 
the octet baryon 
mass in the SU(3) chiral limit (of vanishing $u$-, $d$- and $s$-quark mass) and
we use $M_0=869$ MeV as obtained in a best fit to the data. This value of $M_0$
is consistent with $M_0=(0.77\pm 0.11)$ GeV of ref.\cite{bugra} as found in a 
recent heavy baryon chiral perturbation theory calculation to order $q^4$. The 
third chiral symmetry breaking parameter $b_0$ is then determined through the 
(leading order) mass relation $M_N = M_0 +4m_K^2(b_F-b_D-b_0)-2m_\pi^2(b_0+2
b_F)$ as $b_0= -0.320$ GeV$^{-1}$. Such a  value of $b_0$ leads to a relative
scalar strangeness content in the nucleon of $y=0.1$ and to a pion-nucleon
sigma-term of $\sigma_{\pi N}=31$ MeV.  Whereas $y$ is compatible with the 
present empirical determination $y=0.2\pm0.2$ \cite{gls}, the pion-nucleon 
sigma-term is somewhat too small compared to the empirical value $\sigma_{\pi 
N} = (45\pm 8)$ MeV of \cite{gls}. As mentioned already in \cite{kww}, if one 
stays to linear order in the quark masses, as done here, then both pieces of
information ($y$ and $\sigma_{\pi N}$) cannot be explained by a single value of
$b_0$.  
 
The remaining parameters are adjusted in a best fit to a large amount of 
pion- and photon-induced data. The four range parameters are found as 
\begin{equation} \alpha_{\pi N} = 480\, {\rm MeV}\,, \quad \alpha_{\eta N} =
663\, {\rm MeV}\,, \quad \alpha_{K\Lambda} = 1362\, {\rm MeV}\,, \quad 
\alpha_{K\Sigma} = 785\, {\rm MeV}\,. \end{equation} 
We give here sufficiently many digits to make the numerical results 
reproducible. One observes that the four range parameters are rather close to
the values found in \cite{kww} where only s-wave amplitudes have been
considered, and they are all located in the physically reasonable range
$0.5\dots 1.4$ GeV. At the same time the coefficients of the four
double-derivative s-wave terms  come out as
\begin{equation} d_D= 0.512\,, \quad d_F= -0.509\,, \quad d_0= -0.996\,, 
\quad d_1= 0.173\,, \end{equation}
(in units of GeV$^{-1}$) and those of the four analogous spin-non-flip p-wave
terms as  
\begin{equation} g_D= 0.320\,, \quad g_F= 0.026\,, \quad g_0= -1.492\,, 
\quad g_1= 1.452\,. \end{equation}
Furthermore, the coefficients of the three spin-flip p-wave terms are found as
\begin{equation} h_D= 0.778\,, \quad h_F= -0.029\,, \quad h_1= 0.130\,, 
\end{equation}
(again in units of GeV$^{-1}$). Note that all these coefficients are of
''natural'' size. One also observes that the coefficients $d_{D,F,0,1}$ do not
differ much from those found in our previous work \cite{kww} where only s-wave
amplitudes  have been considered. In essence, the relative weights between the
total cross sections in the different hadronic channels already fix the whole
pattern of s-wave amplitudes and their parameters. The addition of p-waves does
not change this pattern qualitatively. We have furthermore checked that an
equally good description of the $K^-p\to K^-p,\bar K^0n, \pi^0\Lambda, \pi^+
\Sigma^-, \pi^0 \Sigma^0, \pi^+ \Sigma^-$ low-energy data as in ref.\cite{kww}
can be obtained with the (slightly modified) s-wave parameters eq.(27) after
readjusting the range parameters $\alpha_{\bar K N},\alpha_{\pi\Lambda}$ and 
$\alpha_{\pi\Sigma}$.    

Before presenting results for the observables of pion- and photon-induced
$\eta$- and $K$-production we like to mention some generic features of our 
approach. We found that the second order s- and p-wave terms (proportional to 
$b_{D,F,0}, d_{D,F,0,1}, g_{D,F,0,1}, h_{D,F,1}$) are essential in order to 
obtain a good fit of the total cross sections for pion-induced $\eta$- and 
$K$-production. This is quite different from the case of $K^-$-proton elastic
and inelastic scattering, where one can obtain a good description of the low
energy data already with the leading order terms alone in the coupled channel 
scheme \cite{oset}, at the expense of readjusting the pseudoscalar decay 
constant. Furthermore, we have observed that the baryon
anomalous magnetic moments $\kappa_B$ (collected in eq.(3)) are very important
in order to get a good simultaneous reproduction of the $\eta$- and 
$K$-photoproduction total cross sections. Setting all $\kappa_B$ equal to zero 
would lead to a strong reduction of the magnetic p-wave multipoles $M_{1\pm}$
for kaon-photoproduction which however turn out to be important in the energy
region somewhat above threshold. We have also performed a fit taking into 
account the SU(3) symmetry breaking in the pseudoscalar decay constants by 
setting $f=f_\pi=92.4$ MeV for pions, $f=f_K=113.0$ MeV \cite{pdg} for kaons 
and $f=f_\eta=92.6$ MeV \cite{pdg} for the $\eta$-meson. (Note that we are 
identifying throughout the physical $\eta$-meson with the flavor state 
$\eta_8$.). We find that such SU(3)-breaking effects can be balanced by a 
slight change of the coefficients of the second order terms and the range 
parameters. The fit with a single $f=92.4$ MeV is practically indistinguishable
from a fit with three different pseudoscalar decay constants. Therefore we stay
with a single $f=92.4$ MeV as in our previous publications on this and related
subjects. 

We have also performed systematic searches in which the range parameters
$\alpha_n$ (see eq.(9)) entering the p-wave amplitudes were allowed to differ
from those of the s-waves. The resulting improvements were only marginal,
however, so that the introduction of four more parameters did not appear
justified and we stayed with the set eq.(26). Let us now discuss the comparison
with the data in detail. 

Note that a direct comparison of our s- and p-wave coupled channel model with 
elastic $\pi N$-scattering in the energy range of interest here ($1.5$\,GeV\,$
\leq \sqrt{s}\leq 2.0$\,GeV) would not be meaningful since partial waves of
much higher angular momentum and several nucleon resonances contribute in that
channel. Introducing $\pi N$ resonances ($S_{11}(1650)$ etc.) would certainly
improve the situation, but at the expense of additional free parameters. In the
present paper we have ignored such explicit resonance contributions as well as
the coupling to the inelastic $\pi\pi N$ continuum. Deficiencies of the present
results for $\eta N$- and $KY$-production may point to the influence of these 
neglected effects and should be further explored.  

\subsection{Total cross sections for $\pi$-induced $\eta$- and $K$-production}
In Fig.1, we show our results for the total cross sections for pion-induced
$\eta$- and $K$-production: $\pi^-p \to \eta n, K^0\Lambda, K^0\Sigma^0,
K^+\Sigma^-$ and $\pi^+ p \to K^+\Sigma^+$. The data are taken from the
compilation \cite{compil}. The agreement of the coupled channel calculation
including s- and p-wave amplitudes with the existing data is very good, even
for energies considerably above threshold (i.e. up to $p_{\pi,lab} \simeq 2$ 
GeV). The contributions of the s-wave amplitude and the two p-wave amplitudes
to the total cross section are shown separately for each reaction channel. 
With the inclusion of the p-wave one is now able to describe also the data in 
the (pure isospin-3/2) channel $\pi^+ p \to K^+ \Sigma^+$ up to $p_{\pi,lab}=
1.4$ GeV. In this channel the s-wave contributes very little, as already found 
in our previous work \cite{kww}, and the p-wave contribution is completely 
dominant. Furthermore, with the inclusion of p-wave amplitudes one can 
describe the total cross section data for $\pi^- p \to K^+\Sigma^-$ up to 
$p_{\pi,lab}=1.8$ GeV whereas the s-wave approximation starts to break down
around $p_{\pi,lab}=1.2$ GeV. Note that the large $\eta$-production cross 
section  in $\pi^- p \to \eta n$ near threshold (usually interpreted in terms
of the nucleon resonance $S_{11}(1535)$ having a large branching ratio into
$\eta N$) is still dominated by the s-wave amplitude. In the present approach
a s-wave quasi-bound $K\Sigma$-state is formed through the coupled channel 
dynamics. Furthermore, the peak in the $\pi^- p \to K^0\Lambda$ total cross 
section is generated by a strong s-wave cusp effect at the $K\Sigma$-threshold
as already found in our previous work \cite{kww}. As a byproduct of this 
calculation we extract the complex $\eta N$ scattering length $a_{\eta N}= 
(0.32 + 0.25 \,i)$ fm. Whereas its imaginary part agrees with the
recent empirical determination $a_{\eta N}=((0.72\pm0.03)+(0.26\pm0.03)\,i)$ fm
of \cite{batinic}, its real part is about  a factor two too small. As mentioned
in \cite{batinic} the real part Re\,$a_{\eta N}$ cannot exceed 0.4 fm as long
as only one $S_{11}$-resonance is included (as it is the case here) and
therefore the relatively small real part is just consistent with this. 

Altogether it is highly non-trivial to produce the pattern of relative weights 
of s- and p-waves as shown in Fig.1. In a coupled channel calculation all 
reactions get linked together and changes in one channel will immediately 
affect all the others. 
\subsection{Differential cross sections for $\pi$-induced $\eta$- and
$K$-production} 
In Fig.2 we show some typical examples of angular distributions of the
differential cross sections for pion-induced $\eta$- and $K$-production. For
the reaction  
$\pi^- p \to \eta n$ with data taken from \cite{etandif} one finds that the
trend of the data is roughly reproduced, however, with too much deviation from
isotropy. In the case $\pi^- p \to K^+ \Sigma^-$ (data are taken from
\cite{kapsimdif}) one finds a good reproduction of the angular distribution at
low energies. However, moving up in energy the forward-backward asymmetry 
develops in a way opposite to the data. This means that the magnitudes of the 
s- and p-wave amplitudes in this channel are given correctly but their 
interference term has the wrong sign. The same features apply to the channel 
$\pi^- p \to K^0 \Lambda$ where the data stem from \cite{kaplamdif}. It may be
that introduction of the $S_{11}(1650)$-resonance coupling to $K\Lambda$ (and 
possibly to $K\Sigma$) is necessary here. Merely adding a 
$S_{11}(1650)$-resonance contribution to the present coupled channel amplitudes
did not turn out to be successful since in the fit it affects too much the 
s-wave amplitudes of the other (strong and electromagnetic) channels. The 
situation is better for the reaction $\pi^- p \to K^0 \Sigma^0$ as can be seen
in Fig.2. Here the features of the data (taken from \cite{ka0si0dif}) are well
reproduced. Finally we present in Fig.2 an angular distribution for $\pi^+ p
\to K^+ \Sigma^+$ together with the data from \cite{kapsipdif}. For this 
reaction the angular distributions of the existing data show indications of 
structures which go beyond s- and p-waves (i.e. deviations from a parabola in
$\cos\theta$).  Nevertheless the gross features are still reasonably well 
reproduced by our calculation. 

\subsection{Total cross sections for $\eta$- and $K$-photoproduction}
In Fig.3, we show total cross sections for $\eta$-photoproduction,
$\gamma p \to \eta p$, together with the data of \cite{krusche} (MAMI) and
\cite{schoch} (ELSA). As expected the total cross section is dominated here
by the s-wave multipole $E_{0+}$, whose value at threshold is $E_{0+,th}^{(\eta
p)} = (9.8+12.8\,i)\cdot 10^{-3} m_\pi^{-1}$. Again these large cross
sections arise from coupled channel dynamics (forming a $K\Sigma$
quasi-bound state) with no explicit $S_{11}(1535)$ introduced.
Furthermore, we show in Fig.3 the total cross sections for
$K^+$-photoproduction $\gamma p \to K^+\Lambda, K^+\Sigma^0$ together with the
recent data \cite{goers} measured by the SAPHIR collaboration at ELSA. In the
channel $\gamma p \to K^+ \Lambda$ a strong s-wave cusp effect is visible at
the $K\Sigma$-threshold. In fact the data show a structure around $E_\gamma
=1.05$ GeV which is consistent with this interpretation. The (complex) electric
dipole amplitude at threshold has the value $E_{0+,th}^{(K^+\Lambda)} =
(-2.7-3.1\,i)\cdot 10^{-3} m_\pi^{-1}$. Above $E_\gamma=1.2$ GeV
the p-wave multipole amplitudes ($E_{1+}, M_{1\pm}$) become very important in
order to reproduce the energy dependence of the $K^+\Lambda$-data. The same
features apply to the channel $\gamma p \to K^+ \Sigma^0$, where
$E_{0+,th}^{(K^+\Sigma^0)} = (3.8+1.5\,i)\cdot 10^{-3} m_\pi^{-1}$. As in our
previous work \cite{kww} there is a some overestimation  near
threshold in $\gamma p \to K^+\Sigma^0$. In both cases there appears in the
data an enhancement around $E_\gamma \simeq 1.5$ GeV (corresponding to
$\sqrt{s} = 1.9$ GeV). It may be due to a nucleon resonance which couples to
$K\Lambda$ and $K\Sigma$. Our coupled channel calculation can generate only
non-resonant background amplitudes in p-waves and is thus not able to describe
this enhancement if it is due to a resonance. Finally, we show results for
neutral kaon-photoproduction, $\gamma p \to K^0 \Sigma^+$, together with the
recently measured total cross section data from ELSA
\cite{prelim}. The calculation gives an electric dipole amplitude at threshold
of $E_{0+,th}^{(K^0 \Sigma^+)} = (1.6+1.6\,i)\cdot 10^{-3} m_\pi^{-1}$. In this
channel the p-wave multipoles are completely dominant and we find a fair
reproduction of the total cross section data from threshold up to about 
$E_\gamma=1.4$ GeV. Note that in comparison to \cite{mart} our approach does
not have the problem of overpredicting the $\gamma p \to K^0 \Sigma^+$ cross
section. 

Altogether, it is highly non-trivial to generate the pattern of s- and p-wave 
contributions as shown in Fig.3 with its distinct features in each channel,
given all the additional constraints from pion-induced $\eta$- and
$K$-production (Fig.1).      
 
\subsection{Differential cross sections for $\eta$- and $K$-photoproduction}
In Fig.4, we show some typical examples of angular distributions of
differential cross sections for $\eta$- and $K$-photoproduction. In the case
$\gamma p \to p \eta$ the MAMI data \cite{krusche} display an almost isotropic 
angular distribution. The small deviations from isotropy at the higher 
energies $E_\gamma \simeq 0.8$ GeV can be explained in our calculation by the 
interference term of the large s-wave with the small p-wave multipoles. 
However, this good agreement does not necessarily imply that p-waves are
indeed present in $\gamma p \to \eta p $ below $E_\gamma = 0.8 $ GeV since 
higher multipoles (d-waves) could cause a similar effect. The angular 
distributions for $\gamma p \to K^+ \Sigma^0$ of the recent ELSA data 
\cite{goers} are reasonably well reproduced  from threshold up to $E_\gamma 
\simeq 1.7$ GeV in our coupled channel calculation. Similar features apply to
the reaction $\gamma p \to K^+ \Lambda $ for $E_\gamma \leq 1.4$ GeV, although
the measured angular distributions \cite{goers} are (within errorbars) close 
to a straight line (in $\cos\theta$) whereas the calculated ones have larger
curvature ($\cos^2 \theta$-term). Finally, we show in Fig.4 two angular 
distributions of the differential cross section for $\gamma p \to K^0\Sigma^+$
together with the recently measured ELSA-data \cite{prelim}. For photon
energies up to $E_\gamma \simeq 1.4$ GeV the overall behavior of these data 
can be well described by our coupled channel calculation. In summary, we find
in fact that the (recently measured) angular distributions for $\eta$- and 
$K$-photoproduction are better reproduced than those for pion-induced processes
(section 3.2).  

\subsection{Polarization observables for $\eta$- and $K$-photoproduction}
In Fig.5, we show characteristic examples of polarization observables for
$\eta$- and $K$-photopro- duction. The calculated target asymmetry for 
$\gamma p \to \eta p$ is  positive and grows steadily  with  photon energy. 
The existing ELSA data \cite{bock} for $A_t$ are instead oscillating about zero
and do not exceed magnitudes of 0.2. For the photon asymmetry $\Sigma_\gamma$ 
the situation is opposite. The calculation gives  small and negative
values whereas the recent GRAAL measurements \cite{didelez} have found positive
photon asymmetries which grow with energy. Since these features of the
polarization data could in principle be explained by s- and p-wave multipoles,
it appears that the real and imaginary parts of the three p-wave multipoles
($E_{1+}, M_{1\pm}$)  for $\gamma p \to \eta p$ are not given correctly by the
present calculation, whereas their magnitudes are satisfactory.  Unlike the
total and differential cross sections, polarization observables are very
sensitive to the interference pattern of the complex multipole amplitudes. 

Finally, we show  recoil polarizations $\Pi_r$ for $K^+$- and 
$K^0$-photoproduction. In the case $\gamma p \to K^+\Sigma^0$ the trend of the 
recent ELSA data \cite{goers} is roughly reproduced by the calculation. The 
oscillations of $\Pi_r$ for $\gamma p \to K^+\Lambda, K^0 \Sigma^+$ 
\cite{goers,prelim} can, however, not be explained. Clearly, the polarization 
observables cannot yet be satisfactorily understood in our approach which is
strongly constrained by a large amount of additional empirical information. 
This may point to the importance of dynamical effects beyond non-resonant
background amplitudes. To our knowledge there is at present no model (with a
reasonably small number of adjustable parameters) which can simultaneously
explain the huge amount of data considered here.

\section{Summary}
In this work we have extended our non-perturbative chiral SU(3) coupled channel
approach to pion- and photon-induced $\eta$- and $K$-production off protons by
including all strong and electromagnetic p-wave multipoles. The p-wave
amplitudes of the next-to-leading order SU(3) chiral meson-baryon Lagrangian
are identified with a coupled channel potential which is iterated to infinite
orders in a separable Lippmann-Schwinger equation. Seven adjustable parameters 
$(g_{D,F,0,1}, h_{D,F,1})$ enter in the strong interaction p-wave amplitudes. 
The further step to $\eta$- and $K$-photoproduction does not introduce any
additional free parameters since the baryon magnetic moments are well known. 
The latter are important in order to be able to generate large 
$M_{1\pm}$-multipoles for $K$-photoproduction as required by the data. The
drawback of the model at this stage is that the elastic $\pi N$-channel is not
yet handled with sufficient accuracy. 

Altogether, one finds a good reproduction of the total cross sections for
pion- and photon-induced $\eta$- and $K$-production. The gaps left in our
previous work \cite{kww} where only s-waves have been considered are filled by
the p-waves of the coupled channel approach. This requires a highly non-trivial
pattern of relative weights of s- and p-wave contributions. The angular
distributions for $\eta$- and $K$-photoproduction are fairly well 
reproduced throughout. However, there remain some discrepancies in the angular
distributions of pion-induced production and particularly in the polarization
observables of photoproduction. These may point towards the importance of
other dynamics or the need to include further channels (such as the $\eta'$
with its close link to the axial U(1) anomaly in QCD).    
Work along these lines is in progress, together with a more accurate treatment
of the elastic $\pi N$-channel. 
\subsection*{Acknowledgments}
We are grateful J. Didelez and S. Goers for very valuable information on the
experimental data. 

\subsection*{Appendix A}
Here we give the explicit expressions for the p-wave coupling strengths
$C_{ij}^{1\mp}$ in the relevant meson-baryon channels. The channels 
$|\pi N\rangle$, $|\eta N\rangle$, $|K \Lambda \rangle$, $|K \Sigma \rangle$
with total isospin $I=1/2$ and  $|\pi N\rangle$, $|K \Sigma \rangle$ with 
total isospin  $I=3/2$ are labeled with indices $1,\dots, 6$, in this order.

\medskip
\noindent
a) \underline{$p_{1/2}$-channels:}
\begin{eqnarray}
C_{11}^{(1-)} & = & -{1 \over 2 M_0}- {1\over 3}(g_D + g_F + 2 g_0 +4h_D+4h_F)
+2(D+F)^2 P_{\pi\pi}\,, \nonumber \\ C_{12}^{(1-)} & = &-{1\over 3}
(g_D+g_F)+(D+F)(3F-D)P_{\pi\eta}\,,\nonumber \\
C_{13}^{(1-)} & = & {3 \over 8 M_0}+ {1\over 6}(g_D + 3g_F + 2 h_D+6h_F)
-{1\over 4}(D^2+14DF+9F^2) P_{\pi K}\,, \nonumber \\
C_{14}^{(1-)} & = & -{1 \over 8 M_0}+ {1\over 6}(g_D -g_F - 2 g_1 +4h_1 +2h_D
-2 h_F) +{1\over 12}(25D^2 +6DF-39F^2) P_{\pi K}\,, \nonumber \\
C_{22}^{(1-)} & = & {1\over 9}(3g_F -5 g_D -6 g_0) +{1\over 3}(D-3F)^2
P_{\eta\eta}\,, \nonumber \\ 
C_{23}^{(1-)} & = & {3 \over 8 M_0}+ {1\over 18}(12h_1+6h_D+18h_F-g_D -3 g_F -6
 g_1) +{1\over 12}(5D^2+6DF-27F^2) P_{\eta K}\,, \nonumber \\
C_{24}^{(1-)} & = & {3 \over 8 M_0}+ {1\over 6}(g_D - g_F - 6h_D+6h_F)
+{1\over 4}(10DF-D^2-9F^2) P_{\eta K} \,,\nonumber \\
C_{33}^{(1-)} & = & - {1\over 9}(5g_D + 6 g_0 +6h_D)
+{1\over 3}(D^2+3DF+9F^2) P_{KK}\,, \nonumber \\
C_{34}^{(1-)} & = & {1\over 3}(2h_D-g_D)+(3F^2-DF-D^2) P_{KK} \,,\nonumber \\
C_{44}^{(1-)} & = & -{1 \over 2 M_0}+{1\over 3}(2g_F - g_D - 2 g_0 +2h_D-4h_F)
+(2D^2-5DF+2F^2) P_{KK} \,,\nonumber \\
C_{55}^{(1-)} & = & {1 \over 4 M_0}+ {1\over 3}(2h_D+2h_F-g_D - g_F - 2 g_0)
+{1\over 2}(D+F)^2 P_{\pi\pi}\,, \nonumber \\ 
C_{56}^{(1-)} & = & {1 \over 4 M_0}+ {1\over 3}(g_F -g_D-g_1+ 2 h_1-2h_D+2h_F)
+{1\over 6}(3F^2-6DF-D^2) P_{\pi K}\,, \nonumber \\ 
C_{66}^{(1-)} & = & {1 \over 4 M_0}+ {1\over 3}(2h_D+2h_F-g_D - g_F - 2 g_0)
+{1\over 2}(D+F)^2 P_{KK}\,. \end{eqnarray} 

\medskip
\noindent
b) \underline{$p_{3/2}$-channels:}
\begin{eqnarray}
C_{11}^{(1+)} & = & {1\over 3}(2h_D+2h_F -g_D - g_F - 2 g_0)
+{1\over 2}(D+F)^2 P_{\pi\pi} \,, \nonumber \\ C_{12}^{(1+)} & = &-{1\over 3}
(g_D+g_F)+{1\over 2}(D+F)(D-3F)P_{\pi\eta}\nonumber \\ C_{13}^{(1+)} & = & 
{1\over 6}(g_D+3g_F-h_D-3h_F)+D(F-D) P_{\pi K} \,,\nonumber \\
C_{14}^{(1+)} & = & {1\over 6}(g_D -g_F - 2 g_1 -2h_1 -h_D+h_F) +{1\over 3}
(D^2 -3DF+6F^2) P_{\pi K} \,, \nonumber \\
C_{22}^{(1+)} & = & {1\over 9}(3g_F -5 g_D -6 g_0) -{1\over 6}(D-3F)^2
P_{\eta\eta} \,, \nonumber \\ 
C_{23}^{(1+)} & = & - {1\over 18}(6h_1+3h_D+9h_F+g_D +3 g_F +6 g_1)-
D \Big(F+{D\over 3}\Big) P_{\eta K} \,, \nonumber \\
C_{24}^{(1+)} & = & {1\over 6}(g_D -g_F+3h_D-3h_F)+D(F-D)P_{\eta K}\,, 
\nonumber \\ C_{33}^{(1+)} & = & {1\over 9}(3h_D-5g_D - 6 g_0)
-{1\over 6}(D-3F)^2 P_{KK} \,,\nonumber \\
C_{34}^{(1+)} & = & -{1\over 3}(h_D+g_D)+{1\over2}(D+F)(D-3F)P_{KK}\,,
\nonumber \\ C_{44}^{(1+)} & = & {1\over 3}(2g_F - g_D - 2 g_0 -h_D+2h_F)
+{1\over 2}(D+F)^2 P_{KK} \,,\nonumber \\
C_{55}^{(1+)} & = & -{1\over 3}(h_D+h_F+g_D + g_F +2 g_0)-(D+F)^2 P_{\pi\pi} 
\,, \nonumber \\  C_{56}^{(1+)} & = & {1\over 3}(g_F -g_D-g_1- h_1+h_D-h_F)
+{1\over 3}(D^2 +6DF-3F^2) P_{\pi K} \,,\nonumber \\ 
C_{66}^{(1+)} & = & - {1\over 3}(h_D+h_F+g_D + g_F + 2 g_0)
-(D+F)^2 P_{KK} \,. \end{eqnarray} 
The abbreviation $P_{\phi\phi'}$ in eqs.(7,8) stands for  
\begin{equation} P_{\phi \phi'} = {1\over 6E_\phi E_{\phi'}} \biggl( E_\phi + 
E_{\phi'} + { 2E_\phi E_{\phi'}+m_\phi^2 +m_{\phi'}^2 \over 2M_0} \biggr)\,\,.
\end{equation} 
One makes the following observations. The SU(3) Weinberg-Tomozawa meson-baryon
contact vertex contributes only to the $p_{1/2}$-potentials a relativistic
correction proportional to $1/M_0$. The s-channel pole diagrams (i.e. meson 
absorption and subsequent emission on the baryon line) contribute of course 
only to the $p_{1/2}$-potentials, namely with strength $3P_{\phi \phi'}$. The 
contributions of the (crossed) u-channel pole diagrams to the 
$p_{3/2}$-potentials differs by a factor $-2$ from the ones to the 
$p_{1/2}$-potentials (i.e. $-2 P_{\phi \phi'} $ versus $P_{\phi\phi'}$). Note 
that we have used the freedom to add higher order terms in the
small momentum expansion in order to arrive at the fully symmetric expression 
for $P_{\phi \phi'}$.  Finally, there are the p-wave contact terms
in eq.(1) which generate the contributions proportional to $g_{D,F,0,1}$ and
$h_{D,F,1}$.

\begin{figure}
\begin{center}
\epsfig{file=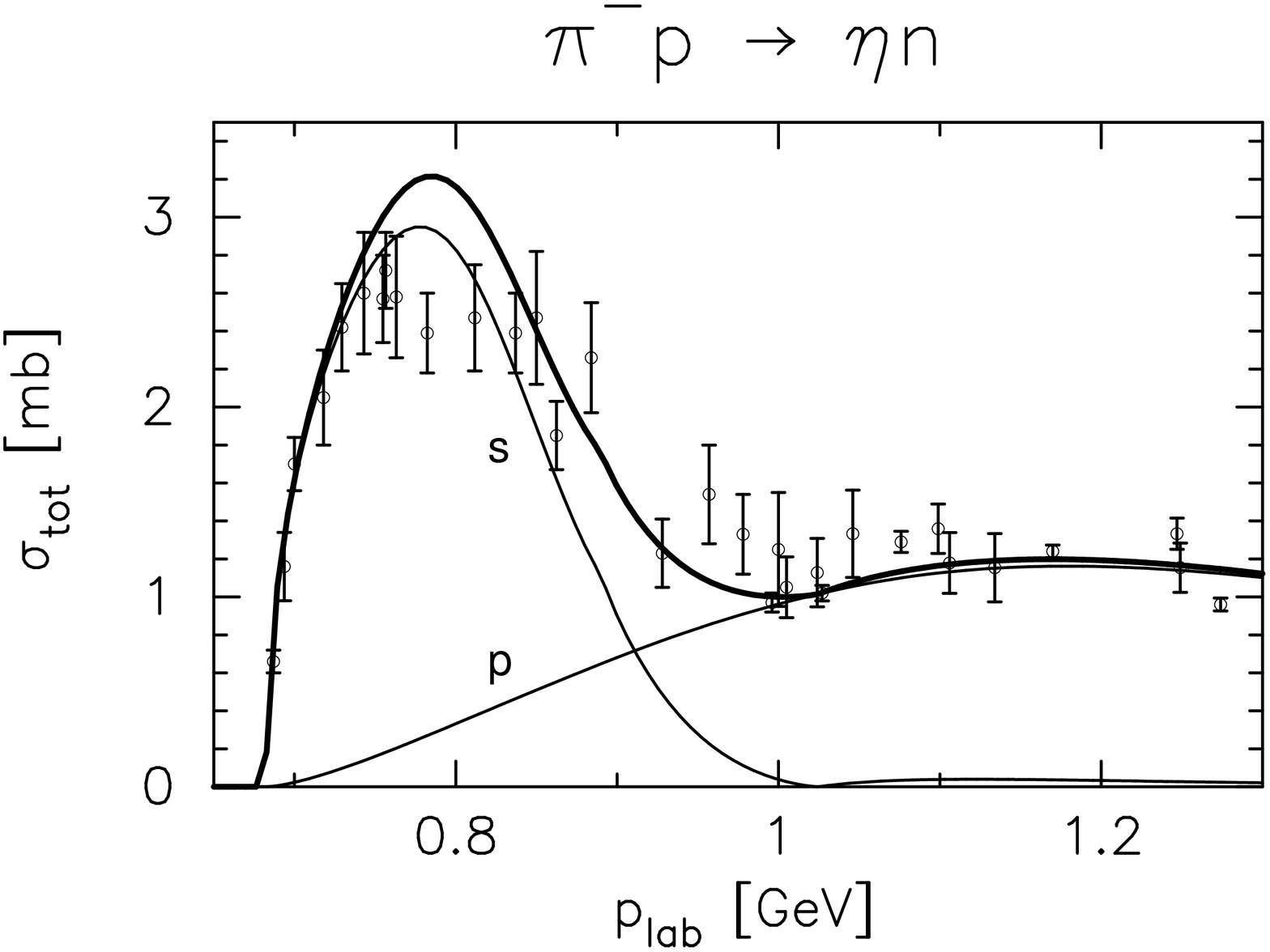,width=60.0mm,angle=-90}
\epsfig{file=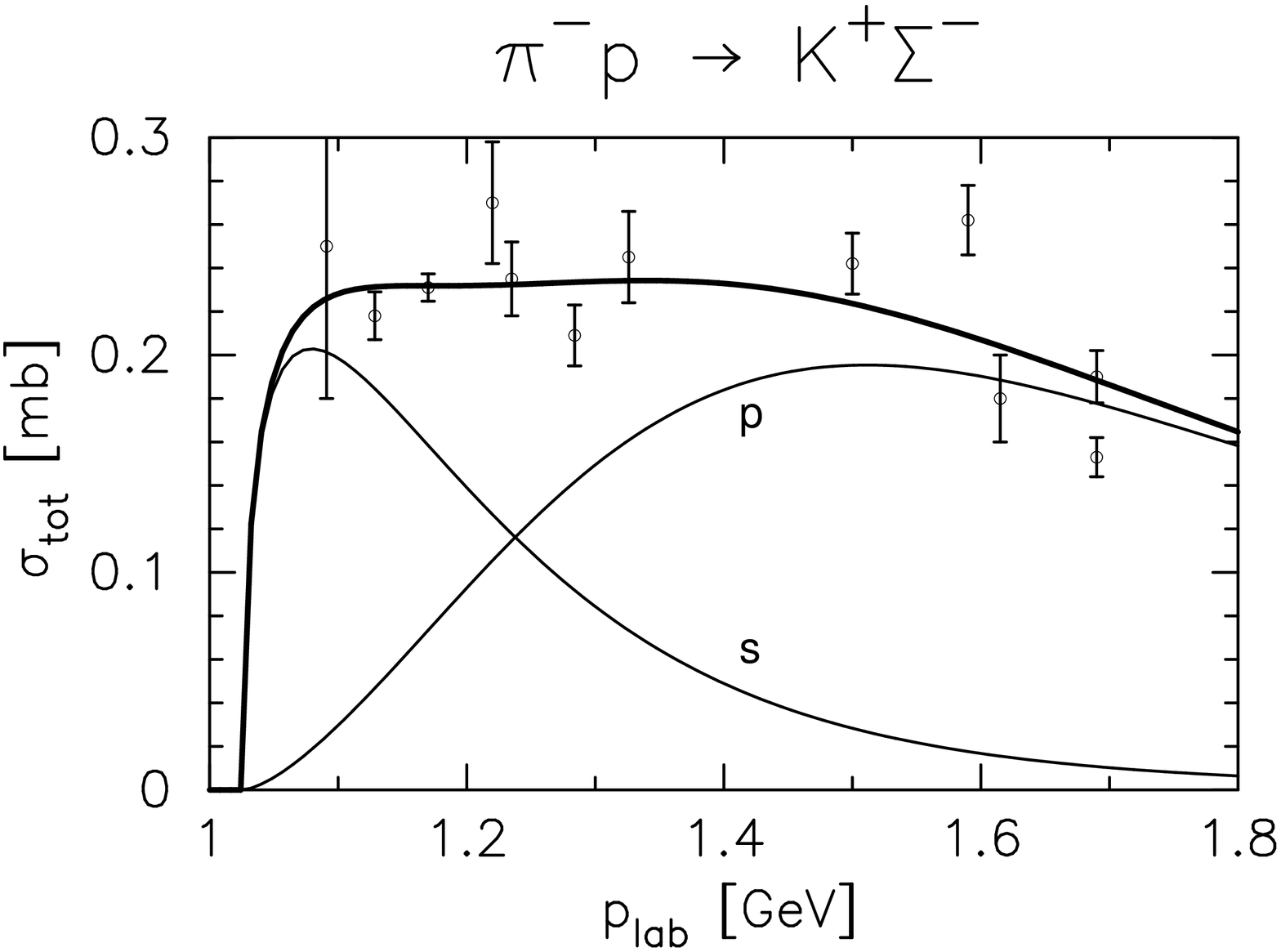,width=60.0mm,angle=-90}
\epsfig{file=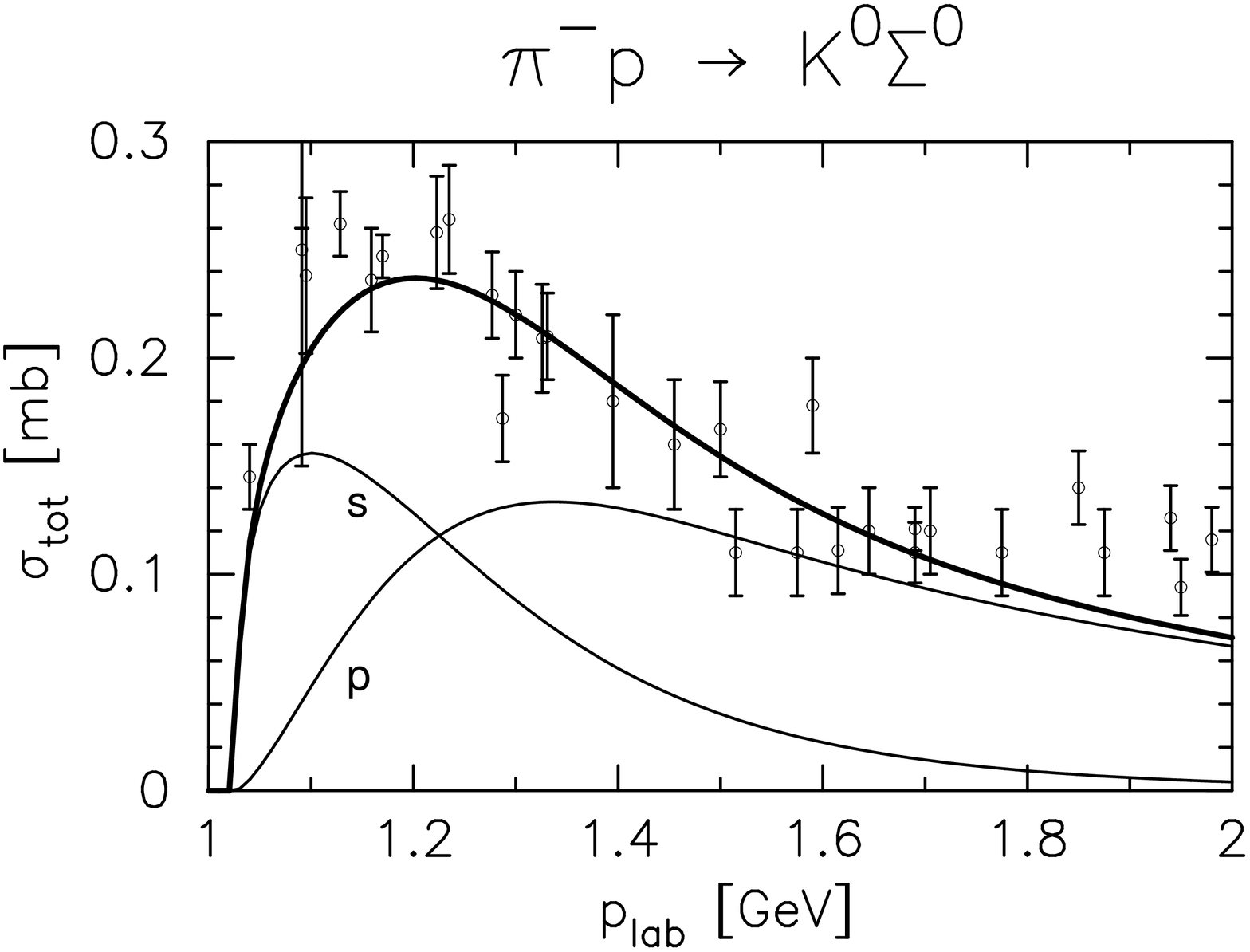,width=60.0mm,angle=-90}
\epsfig{file=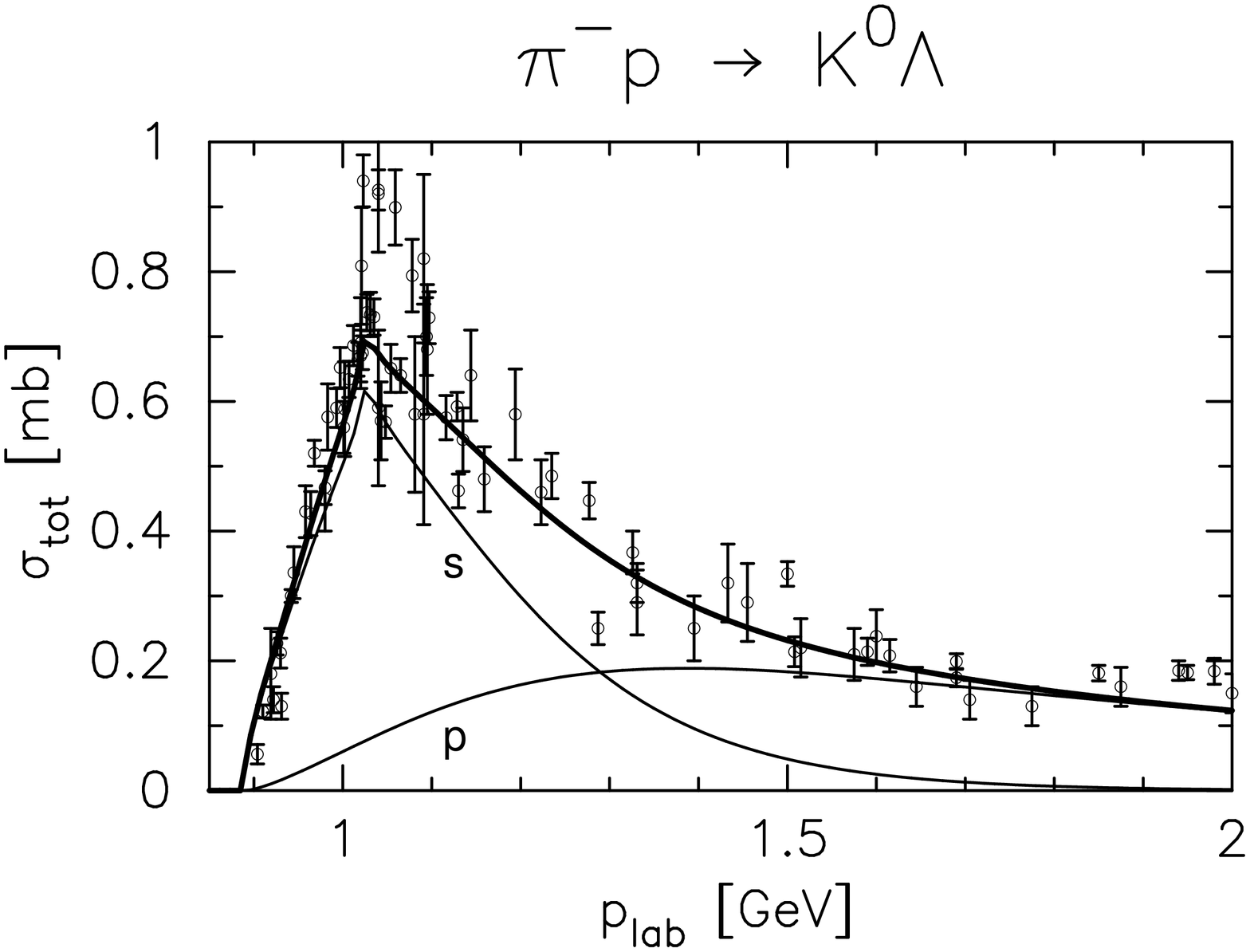,width=60.0mm,angle=-90}
\epsfig{file=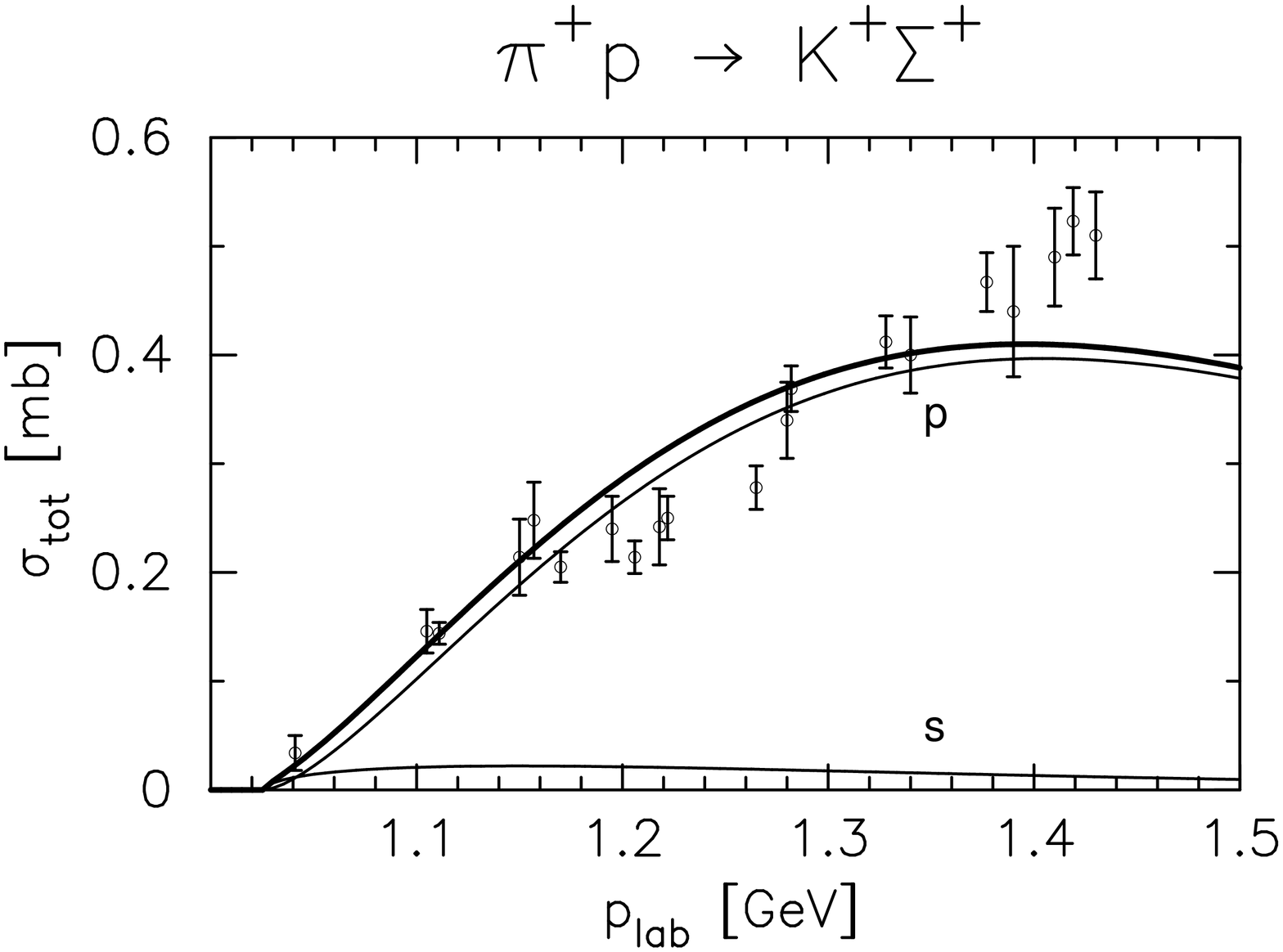,width=60.0mm,angle=-90}
\end{center}
\caption{Total cross sections for pion-induced $\eta$- and
$K$-production as a function of the pion laboratory momentum. The s- and p-wave
contributions are shown separately. The data are taken from the compilation
\cite{compil}. } 
\end{figure}
\newpage
\begin{figure}
\begin{center}
\epsfig{file=d.1.2.ps,width=50.0mm,angle=-90}
\epsfig{file=d.1.3.ps,width=50.0mm,angle=-90}
\epsfig{file=d.2.1.ps,width=50.0mm,angle=-90}
\epsfig{file=d.2.4.ps,width=50.0mm,angle=-90}
\epsfig{file=d.3.2.ps,width=50.0mm,angle=-90}
\epsfig{file=d.3.4.ps,width=50.0mm,angle=-90}
\epsfig{file=d.4.5.ps,width=50.0mm,angle=-90}
\epsfig{file=d.5.1.ps,width=50.0mm,angle=-90}
\end{center}
\caption{Angular distributions of differential cross sections for pion-induced
$\eta$- and $K$-production as a function of $z=\cos\theta$. The data are taken
from \cite{etandif,kaplamdif,ka0si0dif,kapsimdif,kapsipdif}.}  
\end{figure}
\newpage
\begin{figure}
\begin{center}
\epsfig{file=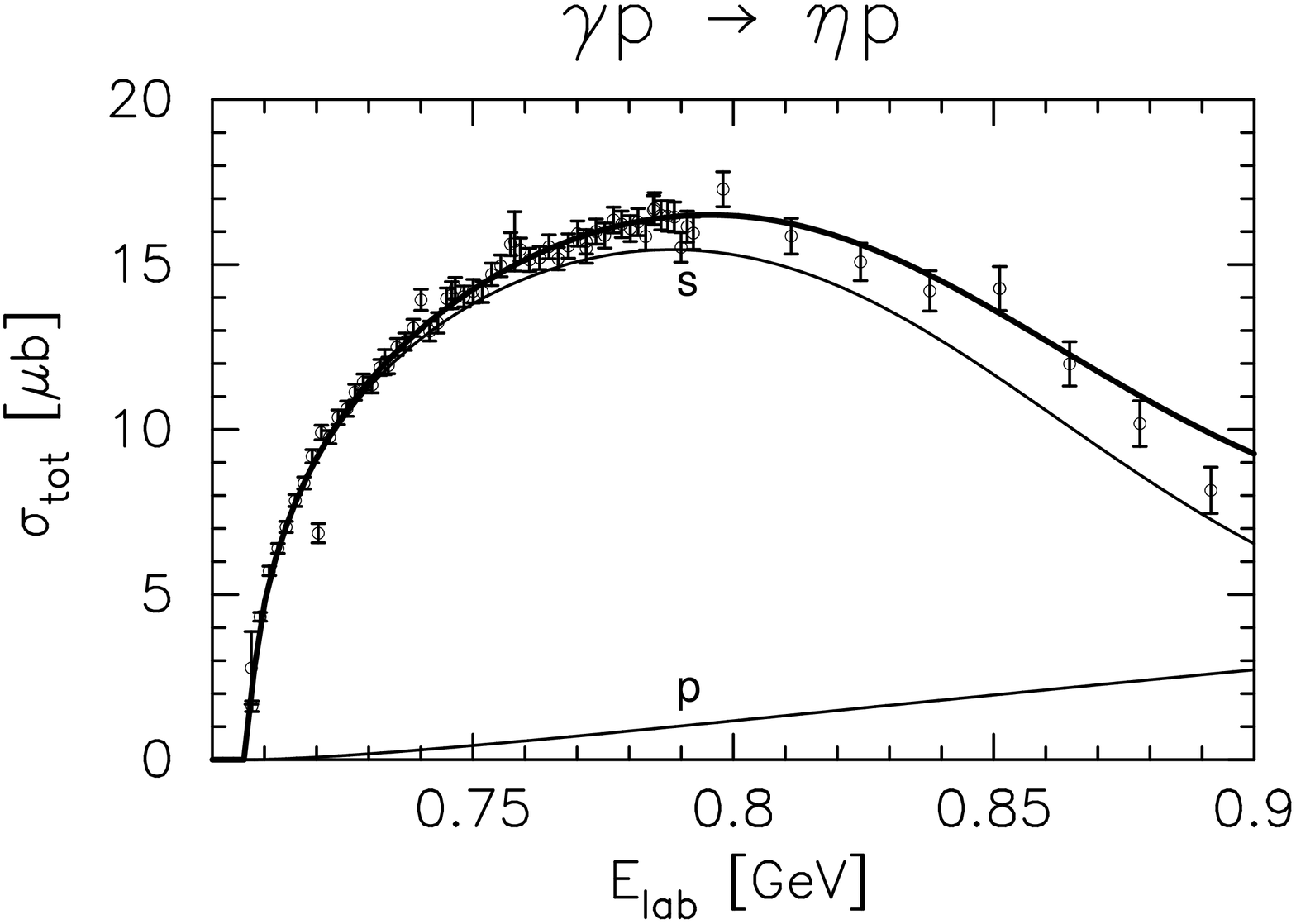,width=55.0mm,angle=-90}
\epsfig{file=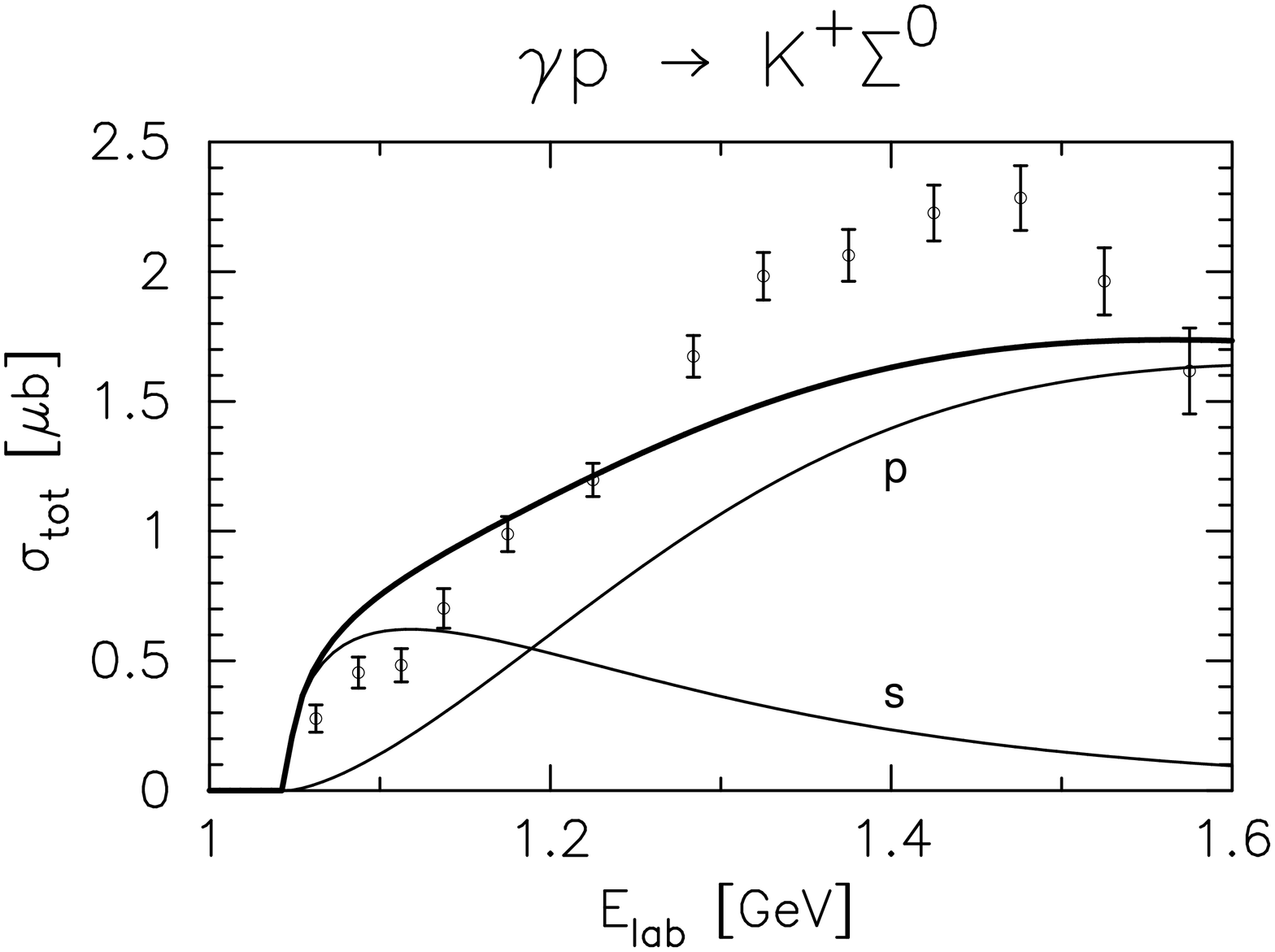,width=55.0mm,angle=-90}
\epsfig{file=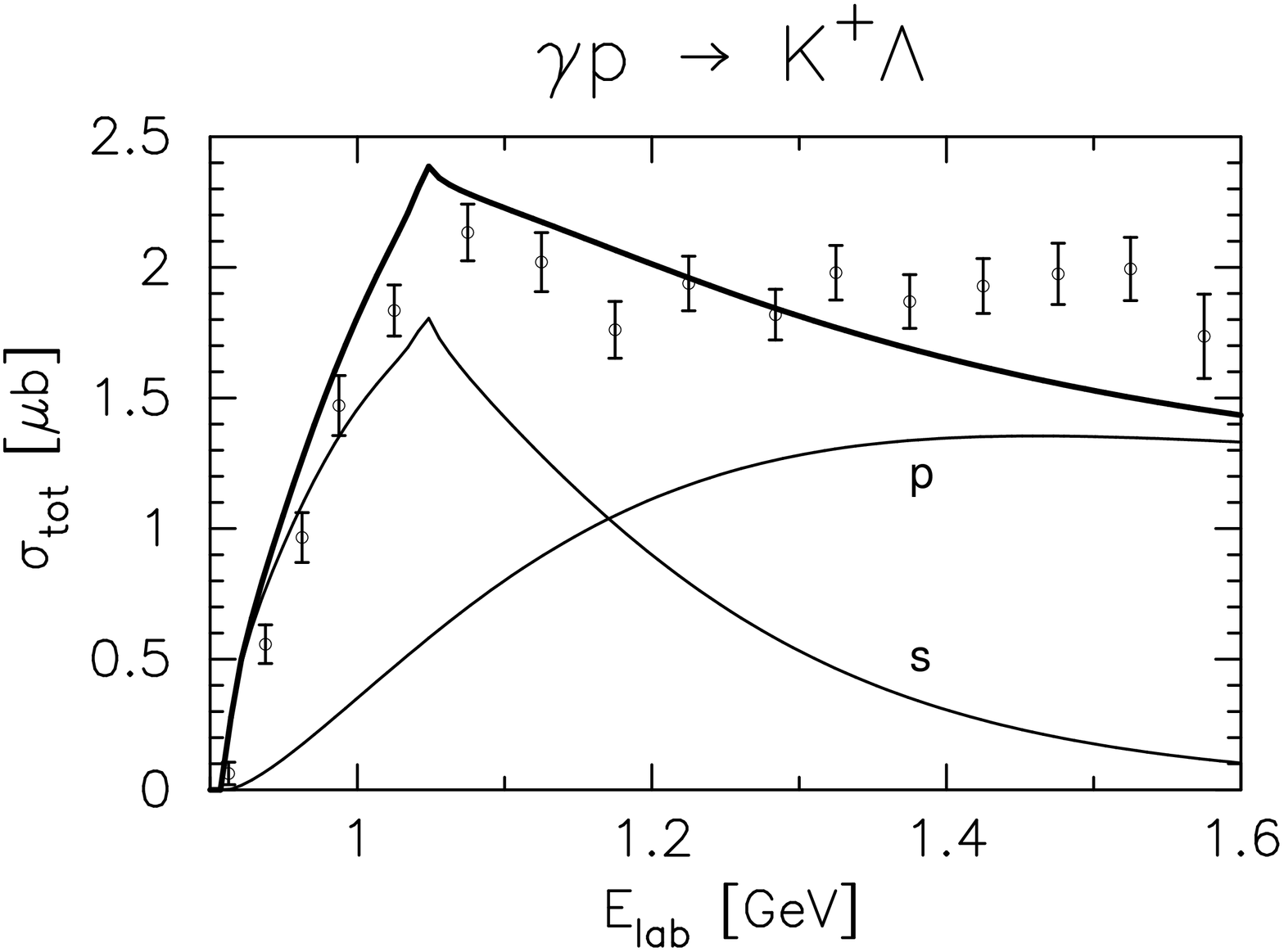,width=55.0mm,angle=-90}
\epsfig{file=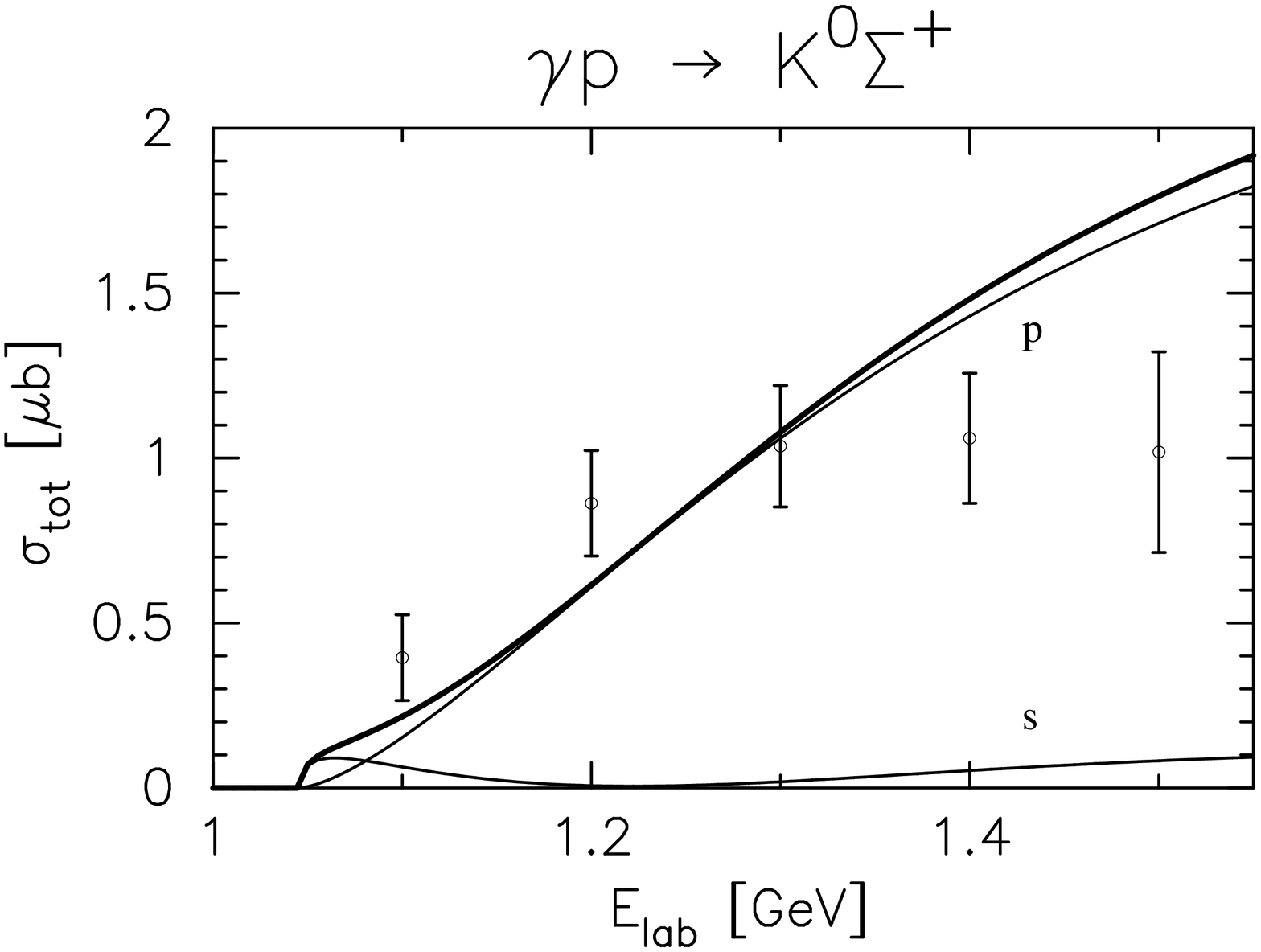,width=55.0mm,angle=-90}
\end{center}
\caption{Total cross sections for $\eta$- and $K$-photoproduction as a function
of the photon laboratory energy. The s- and p-wave contributions are shown
separately. The data are taken from  \cite{krusche,schoch,goers}.}
\end{figure}

\begin{figure}
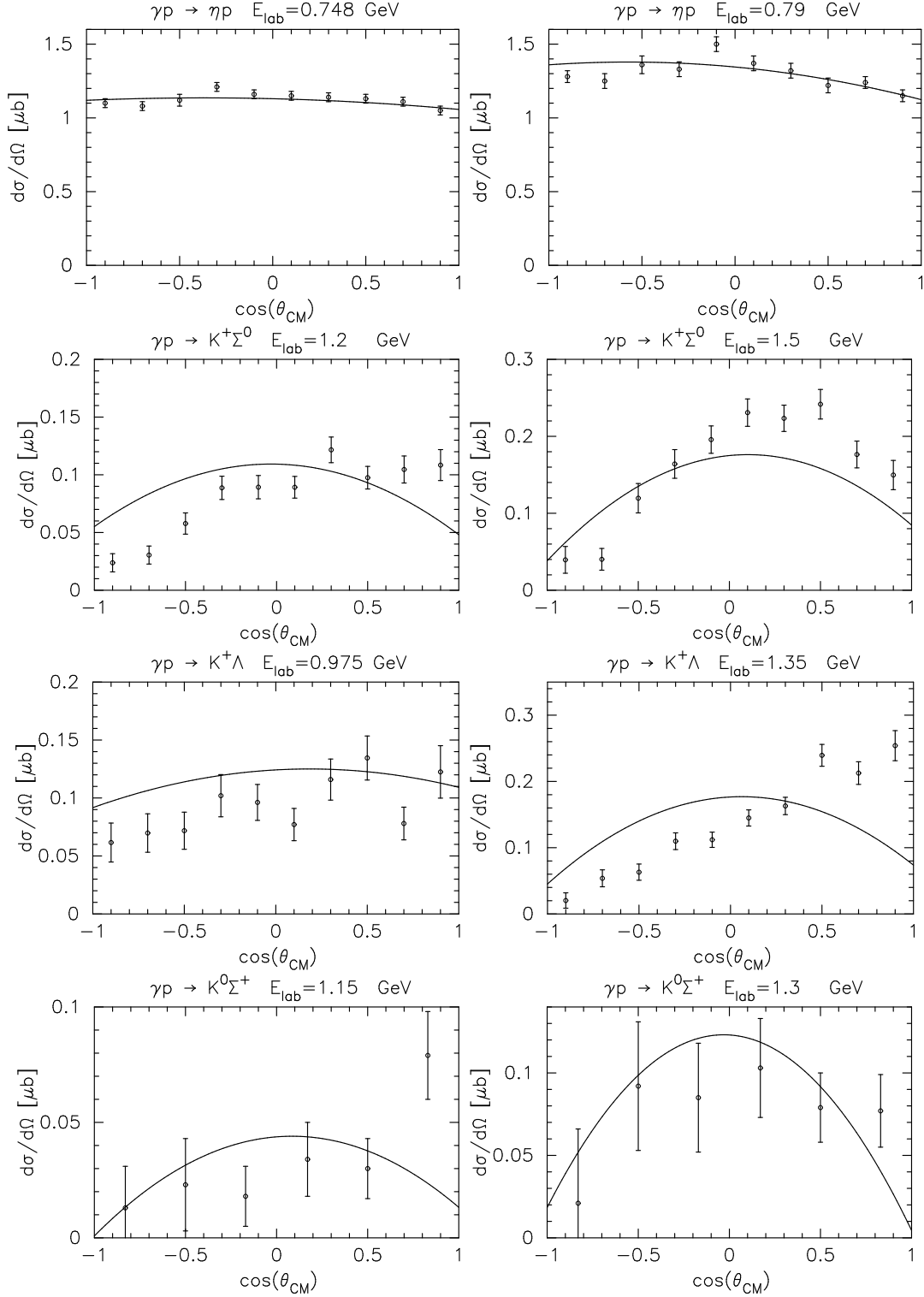

\begin{center}
\epsfig{file=d.6.5.ps,width=50.0mm,angle=-90}
\epsfig{file=d.6.10.ps,width=50.0mm,angle=-90}
\epsfig{file=d.7.2.ps,width=50.0mm,angle=-90}
\epsfig{file=d.7.5.ps,width=50.0mm,angle=-90}
\epsfig{file=d.8.2.ps,width=50.0mm,angle=-90}
\epsfig{file=d.8.6.ps,width=50.0mm,angle=-90}
\epsfig{file=d.9.1.ps,width=50.0mm,angle=-90}
\epsfig{file=d.9.2.ps,width=50.0mm,angle=-90}
\end{center}
\caption{Angular distributions of differential cross sections for $\eta$- and 
$K$-photoproduction as a function of $z=\cos\theta$. The data are taken from 
\cite{krusche,goers}.}
\end{figure}
\newpage

\newpage
\begin{figure}
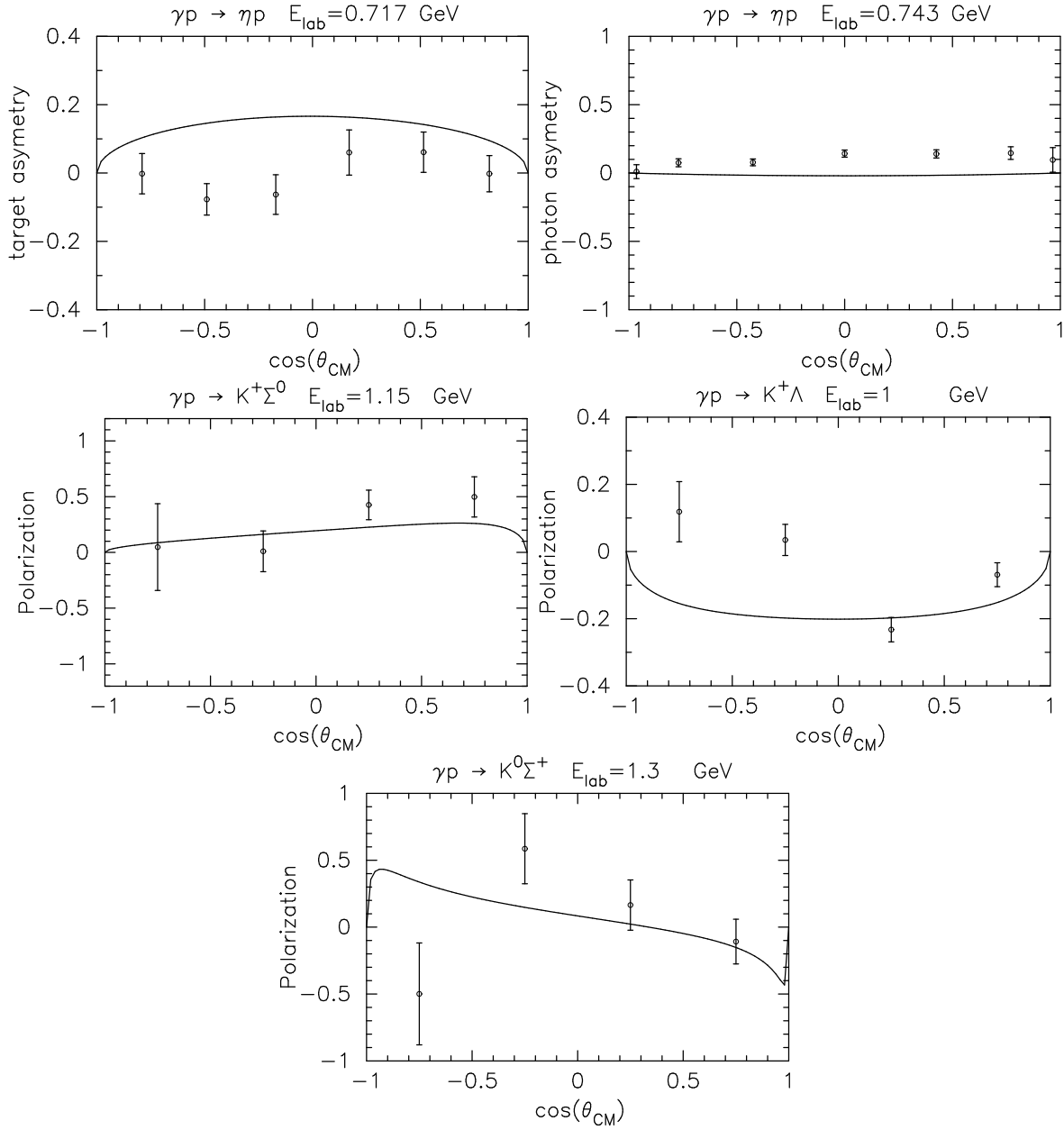

\begin{center}
\epsfig{file=a.1.1.ps,width=55.0mm,angle=-90}
\epsfig{file=b.1.1.ps,width=55.0mm,angle=-90}
\epsfig{file=p.2.1.ps,width=55.0mm,angle=-90}
\epsfig{file=p.3.1.ps,width=55.0mm,angle=-90}
\epsfig{file=p.4.1.ps,width=55.0mm,angle=-90}
\end{center}
\caption{Polarization observables for $\eta$- and $K$-photoproduction as a
function of $z=\cos\theta$. The data are taken from 
\cite{goers,prelim,didelez,bock}.}
\end{figure}

\end{document}